\documentclass[12pt]{article}

\newread\testifexists
\def\GetIfExists #1 {\immediate\openin\testifexists=#1
    \ifeof\testifexists\immediate\closein\testifexists\else
    \immediate\closein\testifexists\input #1\fi}

\usepackage{gthstyle}\usepackage{amsfonts}
\usepackage{amssymb}
\immediate\openin\testifexists=e
    \ifeof\testifexists\immediate\closein\testifexists\else
    \immediate\closein\testifexists\input e\fipsf

\def\Bbb#1{\setbox0=\hbox{$\tt #1$}  \copy0\kern-\wd0\kern .1em\copy0}

\def\bbf#1{\setbox0=\hbox{$#1$} \kern-.025em\copy0\kern-\wd0
        \kern.05em\copy0\kern-\wd0 \kern-.025em\raise.0433em\box0}

\def\d{\delta}        
         \def\k{\kappa}      
                     \def\vv{\varphi}
             
     \def\s{\sigma}

\def\w{\omega}

\def\pa{\partial} \def\ra{\rightarrow}

\def\dd{{\rm d}}  \def\bra{\langle}   \def\ket{\rangle}

\def\fract#1#2{{\textstyle{#1\over#2}}}
\def\ffract#1#2{\raise .3 em\hbox{$\scriptstyle#1$}\kern-.25em/
                \kern-.2em\lower .2 em \hbox{$\scriptstyle#2$}}
\def\fractje#1#2{{\scriptstyle{#1\over#2}}}
\def\half{\fract12}  
\def\halfje{\fractje12}
\def\part#1#2{{\partial#1\over\partial#2}}

\def\iss{\ =\ }

\newcommand{\be}{\begin{eqnarray}}
\newcommand{\ee}{\end{eqnarray}}
\newcommand{\eqn}[1]{(\ref{#1})}

\newcommand{\bi}[1]{\begin{itemize}\item[#1]}

\newcommand{\ei}{\end{itemize}}

\newcommand{\fn}{\footnote}

\newcommand{\newsec}[1]{\section{#1}\setcounter{equation}{0}}

\def\printversion{\setlength{\textheight}{9in}\setlength{\oddsidemargin}{0in}
    \setlength{\textwidth}{6.3in}\setlength{\topmargin}{-0.1in}}

 \newcommand {\eel}[1]{\label{#1}\end{eqnarray}} \newcommand{\crl}[1]{\label{#1}\\}

\printversion 
\begin{document} \begin{titlepage}

\title{\normalsize \hfill ITP-UU-07/39  \\ \hfill SPIN-07/27
\\ \hfill {\tt hep-th/yymmnnn}\\ \vskip 20mm \Large\bf
Emergent Quantum Mechanics and Emergent Symmetries\thanks{Presented at the 13th International Symposium on
Particles, Strings and Cosmology, \emph{PASCOS}, Imperial College, London, July 6, 2007}}

\author{Gerard 't~Hooft}
\date{\normalsize Institute for Theoretical Physics \\
Utrecht University \\ and
\medskip \\ Spinoza Institute \\ Postbox 80.195 \\ 3508 TD
Utrecht, the Netherlands \smallskip \\ e-mail: \tt g.thooft@phys.uu.nl \\ internet: \tt
http://www.phys.uu.nl/\~{}thooft/}

\maketitle

\begin{quotation} \noindent {\large\bf Abstract } \medskip \\
Quantum mechanics is `emergent' if a statistical treatment of large scale phenomena in a locally
deterministic theory requires the use of quantum operators. These quantum operators may allow for symmetry
transformations that are not present in the underlying deterministic system. Such theories allow for a
natural explanation of the existence of gauge equivalence classes (gauge orbits), including the equivalence
classes generated by general coordinate transformations. Thus, local gauge symmetries and general coordinate
invariance could be emergent symmetries, and this might lead to new alleys towards understanding the flatness
problem of the Universe.
\end{quotation}

\vfill \flushleft{\today}

\end{titlepage}

\eject

\newsec{Starting points}\label{intro.sec}
The 19th century view of the physical world was that all dynamical laws of Nature were \emph{deterministic},
which means that every single event in the Universe was assumed to be an irrevocable consequence of
configurations in its past; according to Laplace the fate of every single atom at all times in the future has
been sealed right from the beginning, by the boundary conditions of the Universe at \(t=0\). A theory of the
world would be most satisfactory if it not only uniquely fixes the local dynamical laws, but also provides
for one particular unavoidable choice of the boundary conditions, not only in the spacelike directions, but
also at time \(t=0\), the moment that would now be called the ``Big Bang". This would have been the ultimate
deterministic scenario. Today it is usually dismissed as being totally at odds with what has been learned
from the theory of quantum mechanics. Today, it certainly seems as if the only theories we have are the ones
where the outcomes of most experiments can only be predicted in terms of probabilities. The root of this
situation is the occurrence of a large number of non-commuting operators, describing the values of measurable
quantities.

The observation we wish to investigate further is that non-commuting operators do not necessarily imply a
breakdown of determinism. One may perfectly well introduce non-commuting operators if the evolution law
itself is still deterministic. For instance, in studying the planetary system, it would be totally legal to
introduce an operator that interchanges the positions of Mars and Venus, or an operator that changes the tilt
of the rotation axis of the Earth. Such operators would `violate Bell's inequalities' just like the ones we
use in ordinary quantum mechanics\cite{Bell}\cite{JBU}. Of course, one must ask what the circumstances may be
that would necessitate such a step, in a theory that nevertheless is deterministic. If we study the evolution
of planetary systems or bouncing billiard balls, non-commuting operators never appear to play any substantial
role\fn{There is one important exception: the infinitesimal displacement operators, which are needed to
define, for instance, Lyapunov exponents; see also Section~\ref{math.sec}}. A completely satisfactory answer
to this question has not yet been found, but there do exist systems where one could strongly suspect that
exactly this happens; these theories are \emph{complemented} by the addition of non-commuting operators.

We expect the non-commuting operators to begin playing more essential roles in systems that have a
\emph{dynamical vacuum solution}. This is one dynamical, chaotic solution that may appear to be completely
static at large distance scales, but microscopically it is not static at all. For simplicity, we refer to
this microscopical scale as the `Planck scale'. According to our scenario, it will be fundamentally
impossible to recover deterministic rules of behavior at large scales, just because things are surrounded by
a chaotic vacuum already at the Planck scale (or thereabouts). So we should not be surprised at having an
\emph{effective} theory at large scales that produces statistical fluctuations in its answers to what might
be expected to be observed at large scales. We suspect that the introduction of non-commuting operators may
be essential to understand the statistical features at large scales. This is suspected to be what we call
quantum mechanics today.

It should be emphasized that, in this view, we are not opting for a quantum mechanics that would be only
``approximate". Non-commuting operators evolve quantum mechanically even in deterministic systems (think of
how the Earth-Mars exchange operator would evolve in time). Quantum mechanics as a description of the
evolution laws of operators is probably exactly valid in the same sense that the laws of thermodynamics
follow exactly from classical (or quantum) mechanics at atomic scales. In this work, we furthermore claim
that the mathematical features of quantum mechanics indeed can arise naturally, in particular the need to
introduce wave functions, whose amplitudes squared are to be interpreted as probabilities. There are
problems, however, and it will be important to confront these directly, rather than try to hide them. Thus,
quantum mechanics as it is known today has become the theory enabling us to produce the \emph{best possible
predictions} for the future, given as much information as we can give about the system's past, in any
conceivable experimental setup. Quantum mechanics is \emph{not} a description of the actual course of events
between past and future.

Quantum mechanics will exactly reproduce the statistical features of Nature at a local scale, in our
laboratories. The only effect our present considerations will have on the pursuit of an improved, accurate
theory of quantum gravity and cosmology is that the universe in itself is required to be controlled by
equations beyond quantum mechanics. The argument for this is simple. Quantum mechanics has been tailored by
us to describe the statistical outcomes of experiments when repeated many times, locally in some laboratory.
We may well assume this theory to be exact in describing local statistics. The entire Universe, however (in
particular when we are talking about a closed universe), is itself an `experiment' carried out only once, and
all events in it are unique. The question whether a single event took place or not can only be answered by
`yes' or `no', but there is no probabilistic answer. A theory that yields `maybe' as an answer should be
recognized as an inaccurate theory. If this is what we should believe, then only deterministic theories
describing the entre cosmos should be accepted. There can be no `quantum cosmology'.

New light might be shed precisely on those problems that are connected to our attempts to describe Nature at
the Planck scale, as well as at the cosmological scale. A notorious difficulty is the mystery of the
cosmological constant, which appears to be fantastically small yet not zero.\cite{Paddy} We think that indeed
we have something to say about that. This problem cannot go away as long as we hold on to quantum mechanics
as a primary foundation of our theories rather than an emergent effect.

If the vacuum state indeed is a complicated dynamical solution of local deterministic equations, this implies
among others that two systems can never be completely isolated from one another. They are connected by the
dynamical degrees of freedom of the surrounding vacuum. This should explain why isolated systems appear to
feature stochastic behavior; they simply aren't isolated at all.

We shall furthermore argue that \emph{symmetries} such as rotation symmetry, translation, Lorentz invariance,
but also local gauge symmetries and coordinate reparametrization invariance, might all be \emph{emergent},
together with quantum mechanics itself. They are exact locally, but they are not properties of the underlying
ToE; they certainly are not a property of the boundary conditions of the Universe. We furthermore conjecture
that \emph{information} is not conserved in the deterministic description\cite{GtHdisdet}. This assumption
appears to be necessary in order to obtain a chaotic yet relatively ordered vacuum state, and also was found
to be desired in order to understand the positivity of the hamiltonian.

The notion of `free will' is given a meaning that deviates from most standard views on the
subject;\cite{KochenConway}\cite{RT}\cite{BG} ours\cite{NewScientist} appears to go back as far as Benedict
de Spinoza: an individual's actions are completely dictated by laws of Nature, yet this does not exempt us
from our responsibilities for our actions. The `free will postulate' is presently seen as an axiom in the
reconstruction of the interpretation of quantum mechanics. This postulate will not be needed in the form
usually given. We replace it by the condition that a theory or model of Nature should give appropriate
responses for every conceivable initial state. So, in investigating our model, we have the `freedom' to
choose our initial state at will; whatever that state is, the model should tell me what will happen next.
This is referred to as the `unconstrained initial state' requirement.\cite{GtHfreewill}

\newsec{Mathematics: emergent symmetry on a lattice}\label{math.sec}

Consider a system with deterministic laws\cite{GtHqmdet}\cite{GtHPiombino}: \be {\dd\over\dd t}\vec x(t)=\vec
f(\vec x)\ , \eel{conteom} where \(\vec x\) represents a set of variables such as the positions and the
velocities of the planets at a given time, and \(\vec f\) can be any ordinary vector function. It allows us
to introduce quantum operators such as \be \hat{\vec p}&=&-i{\pa\over\pa\vec x}\ ;\\ \hat H&=&\hat{\vec
p}\cdot\vec f(\vec x)+\vec g(\vec x)\ . \eel{contham} The latter operator is not the standard energy
observable for things such as planets, but the generator of their time evolution: \be {\dd\over\dd t}\vec
x(t)=-i[\,\vec x(t),\,\hat H\,]=\vec f(\vec x)\ . \eel{eolHam}

Thus, we have a description of a deterministic system in a quantum mechanical language. This language allows
us to identify emergent symmetries, but not yet many rigorous examples are known or understood. One
interesting example of an emergent symmetry, or rather the emergent enhancement of a given symmetry, is the
following.\cite{GtHCAsymm} Consider a model of spins on a discrete lattice in one space, one time dimension
(extension of this model to more dimensions is straightforward). The sites on the lattice are indicated by
the integers \((x,\,t)\), where \(t\) is a discretized time variable, but only the \emph{even} sites are
occupied by a single boolean variable \(\s_{x,\,t}\). So, \be\s_{x,\,t}=\pm 1\ ,\qquad x+t\ =\hbox{ even }.
\eel{classlattice} The equation of motion is chosen to be \be \s_{x+1,\,t+1}\iss\s_{x,\,t}\ \s_{x+2,\,t}\
\s_{x+1,\,t-1}\ . \eel{eomlattice} This system has an obvious classical translation symmetry:
\be\pmatrix{x\cr t}\ra\pmatrix{x+\d x\cr t+\d t}\ ;\qquad \d x+\d t\ =\hbox{ even }. \eel{classtrl}

We now claim that the quantum symmetry is larger. To see this, we call \(\s_{x,\,t}=\s^3_{x,\,t}\), and we
introduce the quantum operators \(\s^1_{x,\,t}\) on the odd sites: \(x+t\ =\) odd. On a Cauchy surface, here
defined to be the lines \(t=t_0\) and \(t=t_0-1\), the operator \(\s^1_{x,\,t_0}\) is defined to switch the
variables \(\s^3_{x,\,t-1}\,,\ \s^3_{x-1,\,t}\,,\) and \(\s^3_{x+1,\,t}\), leaving the others fixed. One
quickly establishes how these few switches propagate in time, both to the future and to the past: the
operator \(\s^1_{x,\,t_0}\) switches all variables \(\s^3_{y,\,t}\) for which \(|y-x|<|t-t_0|\).

By inspection then, one finds that the switches produced by four \(\s^1\) operations that are arranged in an
elementary square cancel each other out completely, hence \be \s^1_{x,\,t}\ \s^1_{x-1,\,t+1}\
\s^1_{x+1,\,t-1}\ \s^1_{x,\,t+2}\iss 1\ .\eel{eomswitch} \emph{This is exactly the same equation as the
equation of motion for the \(\s^3\) operators}, eq.~\eqn{eomlattice}! Note, that the \(\s^1\) operators are
only defined on the odd lattice sites. The transformation \be \s^3_{x,\,t}\leftrightarrow\s^1_{x+1,\,t}\ ,
\eel{emergentlattice} is our first example of an emergent symmetry due to the introduction of a
non-com\-mutat\-ive operator such as \(\s^1\). It adds the translations over odd distances in space-time to
the symmetry translations \eqn{classtrl} that we already had.

\newsec{Harmonic oscillators}

One way to continue our discussion is to imagine elementary building blocks for a realistic model. A
realistic model of the universe could consist of a large number of tiny segments that are mutually
interacting with their neighbors. These tiny segments could be discrete, like the lattice elements of the
previous section, but it is better to re-introduce a time continuum. Even if time were to be discrete in some
model, one can still imagine filling up the time segments between the discrete events, even if no events take
place in these continua. One now first formulates how the segments would interact if they were not coupled to
their neighbors, and then adds the interaction, assuming all of these to be deterministic.

If every segment has only a finite number of different states to choose from, then inevitably a
non-interacting segment would evolve in a cyclic motion: it would be periodic. So, prior to the inclusion of
the interaction, our non-interacting universe will consist of many segments moving periodically. Let us first
describe that motion using our new ``quantum" language. Let a segment have a period \(T\), and call the
periodic variable \(\vv\), taking values in the segment \([\,0,\,2\pi)\). Our quantum hamiltonian is taken to
be\fn{the function \(g(\vv)\) in Eq.~\eqn{contham} is taken to be zero here.} \be H=\w p\ ;\qquad
\w={2\pi\over T}\ ;\qquad p=-i{\pa\over\pa\vv}\ . \eel{perham} Because of the periodicity, we must have \be
e^{-iHT}=1\quad \ra\quad H={2\pi n\over T}=\w n\ ;\qquad n=0,\,\pm 1,\,\pm 2,\,\dots\ . \eel{discrham} This
is exactly the spectrum of a quantum harmonic oscillator, except for the emergence of negative energy states.
The states with \(n<0\) are forbidden, for some reason. Of course, we also do not have the `vacuum energy'
\(\half \w\), which would have emerged if we would have required \(e^{-iHT}=-1\). This minus sign is as
harmless as an overall addition of \(\half\w\) to the hamiltonian, but perhaps it will mean something in a
more complete theory; we will ignore it for the time being.

The disappearance of the negative energy modes is very troublesome, however. In a single oscillator, one
might still say that energy is conserved, and once it is chosen to be positive, it will stay positive.
However, when two or more of these systems interact, they might exchange energy, and we will have to explain
why the real world appears to have interactions only in such a way that only positive energy states are
occupied. This is a very difficult problem, and, disguised one way or another, it keeps popping up throughout
our investigations. It still has not been solved in a completely satisfactory manner, but we can try to
handle this difficulty, and one then reaches a number of quite interesting conclusions. In short, our problem
is this: in a deterministic theory, one can reproduce quantum-like mathematics in a multitude of ways, but in
many cases one encounters hamiltonian functions that are either periodic (in case time is taken to be
discrete), or not bounded from below (when time is continuous). Can the real world nevertheless be
approximated by, or rather exactly reproduced in, some deterministic model? What then causes the hamiltonian
of the real world to be bounded below, with a very special lowest energy state, the `vacuum', as a result?
Without this positivity of \(H\), we would not have thermodynamics. The hamiltonian is conjugated to time. Is
there something about time that we are not handling correctly?

To get some ideas, consider two periodic systems, with periods \(T_1=2\pi/\w_1\) and \(T_2=2\pi/\w_2\). It
would be described by the hamiltonian \be H=\w_1\,n_1+\w_2\,n_2\ . \eel{hamtwo} As soon as interactions are
added (which can easily be imagined to be deterministic as well), transition amplitudes will be proportional
to \(\bra n_1,\,n_2|H^\mathrm{int}|m_1,\,m_2\ket\), where \(n_{1,2}\) and \(m_{1,2}\) take the entire range
of integral values between \(-\infty\) and \(\infty\). This would make the state \(|0,\,0\ket\) unstable.
States with \emph{both} \(n_1\) and \(n_2\) negative would not cause a problem; we could interpret these as
the bra states. It is the states \(n_1<0,\ n_2>0\), or vice versa, which must somehow be excluded.

A very crude, but nevertheless suggestive observation is the following. Let us write the time evolution
operator as \(e^{-i(H_1\,t_1+H_2\,t_2)}\). Since \(\vv\) is proportional to time, we could actually replace
\(t_i\) here by \(\vv_i/\w_i\), but the present notation is more suggestive. Then, if \(E_i\) are the
eigenstates of \(H_i\), \be E_1\,t_1+E_2\,t_2 = \half\Big((E_1+E_2)(t_1+t_2)+(E_1-E_2)(t_1-t_2)\Big)\ ,
\eel{diagonalH} so that we see an `uncertainty relation' not only saying that \(E_i\, \d t_i\approx \half\),
but also \be 2\d (t_1+ t_2)\approx 1/(E_1+E_2)\ ,\qquad 2\d(t_1- t_2)\approx 1/|E_1-E_2|\ . \eel{uncertaint}
Since we demand only to have states with \(E_1\,E_2\geq 0\), one easily derives \be
E_1+E_2\geq|E_1-E_2|\quad\ra\quad \d (t_1+ t_2)\leq \d ( t_1- t_2)\ , \eel{timetimeunc} \emph{The spread in
relative time differences must be larger than the spread in average time!} This might indicate that there is
information loss: only the information about average time must play a role in the interactions, whereas
information about relative time, or relative positions, is gradually lost.

\newsec{Information loss} The prototype model with information loss is a simple automaton having four states
\(\{i\},\ i=1,\cdots,4\), with the following evolution rule (here discrete in time): \be
\{1\}\ra\{2\}\ra\{3\}\ra\{1\}\ ;\quad\{4\}\ra\{2\}\ . \eel{infolossrule} In such a simple model, one could
decide once and for all that state \(\{4\}\) is superfluous; after one time step it never is reached anymore.
In a complicate universe, however, it is impossibly difficult to distinguish the states, like state
\(\{4\}\), that can never be reached, from the states that continue to reappear in a periodic motion; what is
more, the universe is too large anyway to ever reach a perfectly periodic mode. Rather than removing the
superfluous states, it is technically more convenient to combine states into equivalence classes. Two states
are equivalent iff within a definite amount of time these two states evolve into exactly the same state.
Thus, states \(\{1\}\) and state \(\{4\}\) form one equivalence class, which we denote as the Dirac ket state
\(|1\ket\). Thus, our model is unitary only in terms of these equivalence classes:
\(|1\ket\ra|2\ket\ra|3\ket\ra|1\ket\).

In our model of two interacting periodic automata, as described in the previous section, we now assume that,
\emph{before} the interaction is switched on, information loss is to be added to the system, such that
information concerning \(\vv_1-\vv_2\) is erased, while information concerning \(\vv_1+\vv_2\) is
preserved\fn{the coefficients of proportionality indeed have to be adapted, see ref\cite{GtHPiombino}.}. Only
then interactions can be admitted. After the interaction, one then again obtains a periodic system, so that
the procedure can be repeated to describe all interactions.

Some of the details of this process have been further described in Ref\cite{GtHPiombino}. One not only sees
the need for information loss to occur in the deterministic system, one also sees that the information loss
will be huge. There are two places in the real world where we expect this information loss to play decisive
roles. One is the physics of black holes. According to the holographic principle, the amount on information
in this universe should grow no faster than one bit of information per Planckian surface element, as in black
holes. The surface area of the universe is huge, so there is a gigantic amount of information that is not
lost, but in Planckian units, the volume of the universe is far bigger, and all bits and bytes in the
3-dimensional bulk apparently do disappear.

The other place where one might suspect information loss to take place is in non-Abelian gauge theories.
indeed, the information equivalence classes do resemble the gauge equivalence classes. In practice, they seem
to play the same role: information concerning the gauge parameters of a theory does not seem to propagate in
time. Rather than saying that the gauge degrees of freedom are ``unphysical", one might conclude that the
gauge parameters contain information that, somewhere during the evolution of our fields, got lost: two states
in a different gauge may evolve both into the same state in the same gauge. Needless to say that this is only
a conjecture, but it seems to make sense.

\newsec{Emergent symmetries}

Thus, unitarity of the quantum evolution only holds in terms of the gauge equivalence classes, as is
routinely taken for granted in gauge theories. Local gauge invariance thus could be interpreted as an
emergent symmetry. What about other symmetries such as translation, rotation, and Lorentz invariance? There
is one easy way to elevate discrete symmetries into continuous ones using quantum operators. Consider for
example a discrete displacement operator \(U\), defined on wave functions \(|\{\psi(x)\}\ket\) by \be
U|\{\psi(x)\}\ket=|\{\psi(x-1)\}\ket\ , \eel{discrdispl} or by the commutation rule \be x\,U=U(x+1)\ .
\eel{discrcomm} Its eigenstates are defined by \be U|p,\,r\ket=e^{ip}|p,\,r\ket\ , \eel{Ueigen} where \(-\pi
< p\leq\pi\), and \(r\) describes other quantum numbers. A fractional displacement operator can then be
defined by \be U(a)=e^{iap}\ ,\eel{fractdispl} where now \(a\) can be any (real or complex) number. Symmetry
under discrete displacements then also implies symmetry under the fractional displacements \eqn{fractdispl},
which is unitary for real \(a\). The only price paid for this extension is that, for fractional \(a\), the
transformation transforms ``ontological" states into quantum superpositions, even if the underlying dynamics
is deterministic.

Thus, our world could be deterministic and discrete at the Planck scale, but since we only know about
continuous symmetries, we are always dealing with quantum superpositions.

If the non-Abelian gauge classes are indeed information equivalence classes, an even more daring conjecture
could be that the same could be true for the classes of the coordinate transformations in general relativity.
Their being continuous could be an emergent phenomenon as described above, and their gauge orbits could be
the information equivalence classes. This could mean that the ``ontological" theory could be defined on an
`ordinary' flat space-time. The nice feature of this conjecture would be that it would explain the strange
preference of our Universe for flat spacial dimensions: dark matter and dark energy cancel out (in the
spacelike direction).

\newsec{A simple model}\label{determinant}

\begin{figure}[ht] \begin{quotation}
 \epsfxsize=100 mm\epsfbox{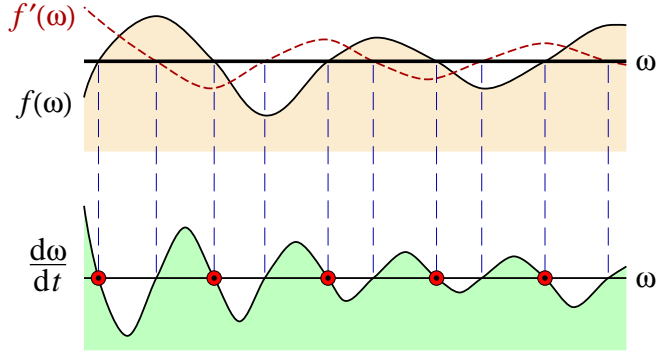}
  \caption{\footnotesize Above: sketches of \(f(\w)\) and of \(f\,'(\w)\) (dotted). Below: a sketch of the
product \(-\k f(\w)\,f\,'(\w)\). The red dots indicate its attractive zeros, which now coincide with the
zeros of \(f\).}
  \label{figure4.fig} \end{quotation}
\end{figure}

In this section, we describe a simple model from which we can derive an existence theorem:
\begin{quotation}\noindent\emph{For any quantum system there exists at least one deterministic model that
reproduces all its dynamics after prequantization}. \end{quotation} To keep the argument transparent, we
consider only a finite dimensional subspace of Hilbert space. Let Schr\"odinger's equation be \be
{\dd\psi\over\dd t}=-iH\psi\ ;\qquad H=\pmatrix {H_{11}&\cdots&H_{1N}\cr \vdots&\ddots&\vdots\cr
H_{N1}&\cdots&H_{NN}}\ . \eel{Schro} We now claim that this system, with the same Hamiltonian, is reproduced
by the following deterministic system. There are two degrees of freedom, \(\w\) and \(\vv\), of which the
latter is periodic: \(\vv\in[0,\,2\pi)\), \ or \ \(\psi(\w,\,\vv)=\psi(\w,\,\vv+2\pi)\). Now, take as
classical equations of motion: \be{\dd\vv(t)\over\dd t}&=&\w(t)\qquad;\crl{dvdt} {\dd\w(t)\over\dd
t}&=&-\k\,f(\w)\,f\,'(\w)\ ,\qquad f(\w)=\det(H-\w)\ . \eel{eomdet} The function \(f(\w)\) has zeros exactly
at the eigenvalues of \(H\). Its derivative, the function \(f\,'(\w)\), has zeros between the zeros of \(f\),
see Fig.~\ref{figure4.fig}.

This way, we achieve the situation that all eigenvalues \(\w_i\ ,\ i=1,\cdots,\,N\) are attractive zeros, as
can be read off from the Figure. Depending on the initial configuration, one of the eigenvalues of the matrix
\(H\) is rapidly approached. If \(\w_{i+\halfje}\) is defined to be the zero of \(f\,'(\w)\) between \(\w_i\)
and \(\w_{i+1}\), then the regions \(\w_{i-\halfje}<\w<\w_{i+\halfje}\) are the equivalence classes
associated to the final state \(\w=\w_i\). Since the convergence towards \(\w_i\) is exponential in time, one
can say that for all practical purposes the limit value \(\w_i\) is soon reached, at which point the system
indeed enters into a limit cycle. At this cycle, the variable \(\vv\) is periodic in time with period
\(T=2\pi/\w\). The pre-quantum states can be Fourier transformed in \(\vv\): \be \psi(\w,\,\vv)=\sum_n
e^{in\vv}\psi_n(\w)\ . \eel{fourier} At \(t\ra\infty\), we have \be
\psi(\w,\,\vv,\,t)\ra\sum_n\psi_n(\w_i,\,0)\,e^{in(\vv-\w_it)}\ . \eel{tinfty}

At this point, we can argue that \(\w_i\) is the ontological energy. In this particular model, we can be even
more explicit. The quantum number \(n\) is absolutely conserved, so, we can use the superselection rule that
the universe settles for one particular value for \(n\), say, \(n=1\). Then, the states \(\psi_1(\w_i)\) can
be exactly identified with the energy eigenstates of the original quantum system. This completes our
mathematical mapping. In this model, interference is a formality: we are always free to consider
probabilistic superpositions, described by superimposing wave functions \(\psi_i\) with different \(i\).

\end{document}